\documentclass[aip]{revtex4-1}

\usepackage{graphicx}

\draft 

\begin{document}

\preprint{}

\title{A sparse coding model with synaptically local plasticity and spiking neurons can account for the diverse shapes of V1 simple cell receptive fields}

\author{Joel Zylberberg}
\email[]{joelz@berkeley.edu}
\affiliation{Department of Physics, University of California, Berkeley CA 94720}
\affiliation{Redwood Center for Theoretical Neuroscience, University of California, Berkeley CA 94720}

\author{Jason Timothy Murphy}
\affiliation{Helen Wills Neuroscience Institute, University of California, Berkeley CA 94720}

\author{Michael Robert DeWeese}
\affiliation{Department of Physics, University of California, Berkeley CA 94720}
\affiliation{Redwood Center for Theoretical Neuroscience, University of California, Berkeley CA 94720}
\affiliation{Helen Wills Neuroscience Institute, University of California, Berkeley CA 94720}

\date{\today}

\begin{abstract}
Sparse coding algorithms trained on natural images can accurately predict the features that excite visual cortical neurons, but it is not known whether such codes can be learned using biologically realistic plasticity rules. We have developed a biophysically motivated spiking network, relying solely on synaptically local information, that can predict the full diversity of V1 simple cell receptive field shapes when trained on natural images. This represents the first demonstration that sparse coding principles, operating within the constraints imposed by cortical architecture, can successfully reproduce these receptive fields. We further prove, mathematically, that sparseness and decorrelation are the key ingredients that allow for synaptically local plasticity rules to optimize a cooperative, linear generative image model formed by the neural representation. Finally, we discuss several interesting emergent properties of our network, with the intent of bridging the gap between theoretical and experimental studies of visual cortex.

\end{abstract}


\maketitle 

\section{Introduction}
A central goal in systems neuroscience is to determine what underlying principles might shape sensory processing in the nervous system. Several coding optimization principles have been proposed, including redundancy reduction~\cite{attneave54,barlow61,atick}, and information maximization~\cite{linsker89,laughlin81,bialekzeeA91,atick92,bialekphysicaA93,bell_sejno,tkacik}, which have both enjoyed some successes in predicting the behavior of real neurons\cite{deweesenetwork96,spikes,atick,dan}. Closely related to these notions of coding efficiency is the principle of sparseness~\cite{foldiak90,olshausen_visres,olshausen96}, which posits that few neurons are active at any given time (population sparseness), or that individual neurons are responsive to few specific stimuli (lifetime sparseness). 

Sparseness is an appealing concept, in part because it provides a simple code for later stages of processing and it is in principle more quickly and easily modifiable by simple learning rules compared with more distributed codes involving many simultaneously active units~\cite{foldiak90,graham2007}. 
There is some experimental evidence for sparse coding in the cortex~\cite{Lennie2003,hromadka,graham2007,vinje,vinje2,haider}, but there are also reports of dense neural activity~\cite{tolhurst} and mixtures of both~\cite{SakataHarris2009} as well. Compounding this, it is not obvious what absolute standard should be used to assess the degree of sparseness in cortex, but it is notable that the relative level of sparseness of cortical responses to natural images increases when a larger fraction of the visual field is covered by the stimulus~\cite{vinje,vinje2,haider}, as a result of inhibitory interneuronal connections~\cite{haider}.  Interestingly, the correlations between the neuronal activities also decreases when a larger area is stimulated, as a result of these inhibitory connections~\cite{haider}. 

In a landmark paper, Olshausen and Field~\cite{olshausen96} reproduced several qualitative features of the receptive fields (RFs) of neurons in primary visual cortex (VI) without imposing any biological constraints other than their hypothesis that cortical representations simultaneously minimize the average activity of the neural population while maximizing fidelity when representing natural images. 
However, agreement with measured V1 simple cell receptive fields was not perfect~\cite{Ringach}. Recently, a more sophisticated version of Olshausen and Field's algorithm~\cite{rehn_sommer} has been developed that is capable of minimizing the number of active neurons rather than minimizing the average activity level across the neural population. This algorithm, called the sparse-set coding (SSC) network~\cite{rehn_sommer}, learns the full set of physiologically observed RF shapes of simple cells in V1, which include small unoriented features, localized oriented features resembling Gabor wavelets, and elongated edge-detectors. We note that, under certain conditions~\cite{donoho} not necessarily satisfied by the natural image coding problem, minimizing the average activity across the neural population ($L_1$-norm minimization), as is done by Olshausen and Field's original Sparsenet algorithm, can be equivalent to minimizing the number of active units ($L_0$-norm minimization), as is achieved by Rehn and Sommer's SSC algorithm.

The SSC model is the only sparse coding algorithm that has been shown to learn, from the statistics of natural scenes alone, RFs that are in quantitative agreement with those observed in V1. It has also been found that sufficiently overcomplete representations (4 times more model neurons than image pixels) that minimize the $L_1$ norm can display the same qualitative variety of RF shapes, but these have not been quantitatively compared with physiologically measured RFs~\cite{warland}. 

Unfortunately, the lack of work on biophysically realistic sparse coding models has left in doubt whether V1 could actually employ a sparse code for natural scenes. Indeed, it is not clear how Olshausen's original algorithm~\cite{olshausen96}, the highly overcomplete $L_1$-norm minimization algorithm~\cite{warland}, or that of Rehn and Sommer~\cite{rehn_sommer}, could be implemented in the cortex.  Rather than  employing local network modification rules such as the synaptic plasticity that is thought to underly learning in cortex~\cite{dayan}, all three of these networks rely on learning rules requiring that each synapse has access to information about the receptive fields of many other, often distant, neurons in the network.

Furthermore, both the SSC sparse coding model that has successfully reproduced the full diversity of V1 simple cell RFs~\cite{rehn_sommer} as well as the $L_1$-norm minimization algorithm that achieved qualitatively similar RFs~\cite{warland} involve non-spiking computational units: continuous-valued information is shared between units while inference is being performed. In cortex, however, information is transferred in discrete, stereotyped pulses of electrical activity called action potentials or spikes. Particularly for a sparse coding model with few or no spikes elicited per stimulus presentation, approximating spike trains with a graded function may not be justified. Spiking image processing networks have been studied~\cite{perrinet,delorme,perrinet_homeo,masquellier,vanrullen, savin}, but none of them have been shown to learn the full diversity of V1 RF shapes using local plasticity rules. It remains to be demonstrated that sparse coding can be achieved within the limitations imposed by biological architecture, and thus that it could potentially be an underlying principle of neural comptutation.

Here we present a biologically-inspired variation on a network originally due to F\"{o}ldi\`{a}k~\cite{foldiak90,falconbridge} that performs sparse coding with spiking neurons. Our model performs learning using only synaptically local rules.  Using the fact that constraints imposed by such mechanisms as homeostasis and lateral inhibition cause the units in the network to remain sparse and independent throughout training, we prove mathematically that it learns to approximate the optimal linear generative model of the input, subject to constraints on the average lifetime firing rates of the units and the temporal correlations between the units' firing rates.  
This is the first demonstration that synaptically local plasticity rules are sufficient to account for the observed diversity of V1 simple cell  RF shapes, and the first rigorous derivation of a relationship between synaptically local network modification rules and the twin properties of sparseness and decorrelation.

Finally, we describe several emergent properties of our image coding network, in order to elucidate some experimentally testable hallmarks of our model. Interestingly, we observe a lognormal distribution of inhibitory connection strengths between the units in our model, when it is trained on natural images; such a distribution has previously been observed in the excitatory connections between neurons in rat V1~\cite{song}, but the inhibitory connection strength distribution remains unknown. 

\section{Results}

\subsection*{Our Sparse And Independent Local network (SAILnet) learns receptive fields that closely resemble those of V1 simple cells}

Our primary goal is to develop a biophysically inspired network of spiking neurons that learns to sparsely encode natural images, while employing only synaptically local learning rules.
Towards this end, we implement a network of spiking, leaky integrate-and-fire units~\cite{dayan} as model neurons. As in many previous models~\cite{rozell,perrinet,falconbridge,foldiak90,hopfield84}, each unit has a time dependent internal variable $u_i(t)$ and an output $y_i(t)$ associated with it. The simulation of our network operates in discrete time. The neuronal output at time $t$, $y_i(t)$, is binary-valued: it is either 1 (spike) or 0 (no spike), whereas the internal variable $u_i(t)$ is a continuous-valued function of time that is analogous to the membrane potential of a neuron. When this internal variable exceeds a threshold $\theta_i$, the unit fires a punctate spike of output activity that lasts for one time step. This thresholding feature plays the role of neuronal voltage-gated ion channels (represented, as in Hopfield's~\cite{hopfield84} circuit model, by a diode) whose opening allows cortical neurons to fire. Other units in the network, and the inputs $X_k$, which are pixel intensities in an image, modify the internal variable $u_i(t)$ by injecting current into the model neuron. The structure of our network, and circuit diagram of our neuron model, are illustrated in Fig.~1. The dynamics of SAILnet neurons are discussed in detail in the Methods section.

We assess the computational output of each neuron in response to a stimulus image $X$ by counting the number of spikes emitted by that neuron, $n_i = \sum_t y_i(t)$, following stimulus onset for a brief period of time lasting five times the time constant $\tau_{RC}$ of the RC circuit.  Our simulation updates the membrane potential every $0.1~\tau_{RC}$ , thus there are 50 steps in the numerical integration following each stimulus presentation. Consequently, at least in principle, 50 is the maximum number of spikes we could observe from one neuron in response to any image. We note that one could instead use first-spike latencies to measure the computational output~\cite{vanrullen,delorme}; these two measures are highly correlated in our network, with shorter latencies corresponding to greater spike counts (data not shown). The network learns via rules similar to those of  F\"{o}ldi\'{a}k~\cite{foldiak90,falconbridge}. These rules drive each unit to be active for only a small but non-zero fraction of the time (lifetime sparseness) and to maintain uncorrelated activity with respect to all other units in the network:

\begin{eqnarray}
\label{eq:learning}
\Delta W_{i m} &=& \alpha ( n_i n_m - p^2) \nonumber \\
\Delta Q_{i k} & = & \beta n_i (X_k - n_i Q_{i k}) \\
\Delta \theta_i &=& \gamma(n_i - p), \nonumber 
\end{eqnarray}

where $p$ is the target average value for the number of spikes per image, which defines each neuron's lifetime sparseness, and $\alpha$, $\beta$, and $\gamma$ are learning rates --- small positive constants that determine how quickly the network modifies itself. Updating the feed-forward weights $Q_{i k}$ in our model is achieved with Oja's implementation~\cite{Oja} of Hebb's rule; this rule is what drives the network to represent the input. Note that because the firing rates are low ($p=0.05$ spikes per image, for the results shown in this paper), and spikes can only be emitted in integer units, our model implicitly allows only small numbers of neurons to be active at any given time (so called ``hard" sparseness, or $L_0$ sparseness), similar to what is achieved by other means in some recent non-spiking sparse coding models~\cite{rehn_sommer,rozell}.

These learning rules can be viewed as an approximate stochastic gradient descent approach to the \emph{constrained} optimization problem in which the network seeks to minimize the error between the input pixel values $\{X_k\}$, and a linear generative model formed by all of the neurons $\overline{X_k} = \sum_i n_i Q_{ik}$, while maintaining fixed average firing rates and no firing rate correlations. This constrained optimization interpretation of our learning rules, and the approximations involved, are discussed in the Methods section. 

In Fig.~2, we demonstrate that the activity of the SAILnet units can be linearly decoded to recover (approximately) the input stimulus. The success of linear decoding in a model that encodes stimuli in a non-linear fashion is a product of our learning rules, and it has been observed in multiple sensory systems~\cite{bialek} and spiking neuron models optimized to maximize information transmission~\cite{bialekphysicaA93,deweesenetwork96}.

Our learning rules encourage all neurons to have the same average firing rate of $p$ spikes per image, which may at first appear to be at odds with the observation~\cite{hromadka} that cortical neurons display a broad distribution of activities --- firing rates vary from neuron to neuron.

However, when trained on natural images, neurons in SAILnet can actually exhibit a fairly broad range of firing rates. Moreover, the mean firing rate distribution ranges from
approximately lognormal to exponential in response to natural image stimuli, depending on the mean contrast of the stimulus ensemble with which they are probed. We discuss this further in the Firing Rates section below. 

We emphasize here that each of our learning rules is ``synaptically" local: the information required to determine the change in the connection strength at any synaptic junction between two units is merely the activity of the pre- and post-synaptic units.  The inhibitory lateral connection strengths, for example, are modified according to how many spikes arrived at the synapse, and how many times the post-synaptic unit spiked. The information required for the unit to modify its firing threshold is the unit's own firing rate. Finally, the rule for modifying the feed-forward connections requires only the pre-synaptic activity $X_k$, the post-synaptic activity $n_i$, and the present strength of that connection $Q_{ik}$. This locality is a desirable model feature because learning in cortex is thought~\cite{dayan} to occur by the modification of synaptic strengths and thus by necessity should depend only upon information available locally at the synapse. 

By contrast, much previous work~\cite{olshausen96,warland,rehn_sommer,perrinet_homeo} has used a different learning rule for the feed forward weights: $\Delta Q_{ik} \propto n_i(X_k -\sum_j n_j Q_{jk})$.
This rule is non-local because the update for the connection strength between input pixel $k$ and unit $i$ requires information about the activities and feed-forward weights of every other unit in the network (indexed by $j$). It is unlikely that such information is available to individual synapses in cortex. Interestingly, in the limit of highly sparse and uncorrelated neuronal activities, our local learning rule approximates the non-local rule used by previous workers~\cite{olshausen96,rehn_sommer,perrinet_homeo}, when averaged over several input images; we provide a mathematical derivation of this result in the Methods section. This suggests an additional reason why sparseness is beneficial for cortical networks, in which plasticity is local, but cooperative representations may be desired.   

We trained a 1536-unit SAILnet with sparseness $p=0.05$ on $16 \times 16$ pixel image patches drawn randomly from whitened natural images from the image set of Olshausen and Field~\cite{olshausen96}.  The network is six-times
overcomplete with respect to the number of input pixels.  This mimics the anatomical fact that V1 contains many more neurons than does LGN, from which it receives its inputs. Owing to the computational complexity of the problem --- there are $\mathcal{O}(N^2)$ parameters to be learned in a SAILnet model containing $N$ neurons --- we found it prohibitive to consider networks that are much more than $6 \times$ overcomplete.

Our 
six-times overcompleteness is in a sense analogous to the  
three-times overcompleteness of the SSC network described by Rehn and Sommer~\cite{rehn_sommer}, since the outputs of their computational units could be either positive or negative, while our model neurons can output only one type of spike. Thus, each of their units can be thought of as representing a pair of our neurons, with opposite-signed receptive fields.

The RFs of 196 randomly selected units from our SAILnet are shown in Fig.~3, as measured by their spike-triggered average activity in response to whitened natural images. These are virtually identical to the feed-forward weights of the units; in the Methods section, we discuss why this must be the case.

To facilitate a comparison between the SAILnet RFs, and those measured in macaque V1 (courtesy of D. Ringach), we fit both the SAILnet, and the macaque RFs to Gabor functions. As in the SSC study of Rehn and Sommer~\cite{rehn_sommer}, only those RFs that could be sensibly described by a Gabor function were included in Fig.~3; for example, we excluded RFs with substantial support along the square boundary, suggesting that the RF is only partly visible. In the Methods section, we discuss the Gabor fitting routine and the quality control measures we used to define and identify meaningful fits.

Our SAILnet model RFs show the same diversity of shapes observed in macaque V1, and in the non-local SSC model~\cite{rehn_sommer}. They consist of three qualitatively distinct classes of neuronal RFs: small unoriented features, localized and oriented Gabor-like filters, and elongated edge-detectors. Our SAILnet learning rules approximately minimize the same cost function as the SSC model~\cite{rehn_sommer}, albeit with constraints as opposed to unconstrained optimization, which explains how it is possible for SAILnet to learn similar RFs using only local rules. Furthermore, in our model, the number of co-active units is small, owing to the low average lifetime neuronal firing rates, and the fact that spikes can only be emitted in integer numbers. This feature is similar to the L$_0$-norm minimization used in the SSC model of Rehn and Sommer~\cite{rehn_sommer} and the LCA model of Rozell and colleagues~\cite{rozell}.

This is the first demonstration that a network of spiking neurons using only synaptically local plasticity rules applied to natural images can account for the observed diversity of V1 simple cell RF shapes.

\subsection*{SAILnet units exhibit a broad distribution of mean firing rates in response to natural images}
 
Our learning rules (Eq.~\ref{eq:learning}) encourage every unit to have the same target value, $p$, for its average firing rate, which might appear to be inconsistent with observations~\cite{hromadka,baddeley,abeles} that cortical neurons exhibit a broad distribution of mean firing rates. However, we find that SAILnet, too, can display a wide range of mean rates, as we now describe.

To determine the distribution of mean firing rates across the population of model neurons in our network, we first trained a 1536-unit SAILnet on $16 \times 16$ pixel patches drawn from whitened natural images, and then presented the network with $50,000$ patches taken from the training ensemble. Our measurement was performed with all learning rates set to zero, so that we were probing the properties of the network at one fixed set of learned parameter values, rather than observing changes in network properties over time. 

We then counted the number of spikes per image from each unit to estimate each neuron's average firing rate, as it might be measured in a physiology experiment. 
The distribution of these mean firing rates is fairly broad and well-described by a lognormal distribution (Fig.~4a). This distribution is strongly non-monotonic, clearly indicating that it is poorly fit by an exponential function.

Subsequently, we probed the same network (still with the learning turned off, so that the network parameters were identical in both cases) with $50,000$ low-contrast images consisting of patches from our training ensemble with all pixel values multiplied by $1/3$. We found that the firing rate distribution was markedly different than what we found when the network was probed with higher-contrast stimuli. In particular, it became a monotonic decreasing function that was similarly well-described by either a lognormal or an exponential function (Fig.~4b).

From the dynamics of our leaky integrate-and-fire units, it is clear that the low contrast stimuli with reduced pixel values will cause the units to charge up more slowly and subsequently to spike less in the allotted time the network is given to view each image. Consequently, the firing rate distribution gets shifted towards lower firing rates. However, negative firing rates are impossible, so in addition to being shifted, the low-firing-rate tail of the distribution is effectively truncated. Note that truncating the lognormal distribution anywhere to the right of the peak results in a distribution that looks qualitatively similar to an exponential.

Mean firing rates in primary auditory cortex (A1) have been reported by one group~\cite{hromadka} to obey a lognormal distribution, whether spontaneous or stimulus-evoked in both awake and anesthetized animals.  However, exponentially distributed spontaneous mean firing rates have also been reported in awake rat A1~\cite{gaese}. 
Although several groups have measured the distribution of firing rates over time for individual neurons~\cite{baddeley, abeles}, we are unaware of a published claim regarding the distribution of mean firing rates in visual cortex.

Recall that our learning rules encourage the neurons to all have the same average firing rate. This fact may be puzzling at first given the spread in mean firing rates apparent in the distributions shown in Fig.~4. There are two main effects to consider when making sense of this: finite measurement time, and non-zero step-sizes for plasticity.

The first effect relates to the fact that there is intrinsic randomness in the measurement process --- which randomly selected image patches happen to fall in the ensemble of probe stimuli --- so that the measured distribution tends to be broader than the ``true" underlying distribution of the system. To check that this effect is not responsible for the broad distribution in firing rates, we computed the variance in the measured firing rate distribution after different numbers of images were presented to the network. The variance decreased until it reached an asymptotic value after approximately $25,000$--$30,000$ image presentations (data not shown). Thus, the $50,000$ image sample size in our experiment is large enough to see the true distribution; finite sample-size effects do not affect the distributions that we observed. 

The other, more interesting, effect that gives rise to a broad distribution of firing rates is related to learning. While the network is being trained, the feed-forward weights, inhibitory lateral connections, and firing thresholds get modified in discrete jumps, after every image presentation (or every batch of images, see the Methods section for details). Since those jumps are of a non-zero size -- as determined by the learning rates $\alpha$, $\beta$, and $\gamma$  -- there will be times when the firing threshold gets pushed below the specific value that would lead to the unit having exactly the target firing rate, and the unit will thus spike more than the target rate. Similarly, some jumps will push the threshold above that specific value, and the unit will fire less than the target amount. Even after learning has converged, and the parameters are no longer changing \emph{on average} in response to additional image presentations, the network parameters are still bouncing around their average (optimal) values; any image presentation that makes a neuron spike more than the target amount results in an increased firing threshold, while any image that makes the neuron fire less than the target amount leads to a decreased firing threshold. Recent results~\cite{clopath} suggest that the sizes of these updates (jumps) are quite large for real neurons. Interestingly, this indicates that the observed broad distributions in firing rate~\cite{hromadka} do not rule out the possibility that homeostatic mechanisms are driving each neuron to have the same average firing rate. 

Reducing the SAILnet learning rates $\alpha$, $\beta$ and $\gamma$ does reduce the variance of the firing rate distributions, but our qualitative conclusions --- non-monotonic, approximately lognormal firing rate distribution in response to images from the training set, and monotonic, exponential/lognormal distribution in response to low contrast images --- are unchanged when we use different learning rates for the network (data not shown).

\subsection*{ Pairs of SAILnet units have small firing rate correlations. }

Recent experimental work~\cite{ecker, renart} has shown that neurons in visual cortex tend to have small correlations between their firing rates. In order to facilitate a comparison between our model, and the physiological observations, we have measured the (Pearson's)  linear correlation coefficients between spike counts of SAILnet units, in response to an ensemble 30,000 natural images. These correlations (Fig.~5) tend to be near zero, as is observed experimentally~\cite{ecker}, while the experimental data show a larger variance in the distribution of correlation coefficients than we observe with SAILnet. We note that, like the firing rate distribution (discussed above), the distribution of correlation coefficients is affected by the update sizes (learning rates) in the simulation, with larger update sizes leading to a larger variance of the measured distribution.

In Fig.~5, the distribution appears truncated on the left. This effect arises because there is a lower bound on the correlation between the neuronal firing rates that arises when the two neurons are \emph{never} co-active. The low mean firing rate of $p = 0.05$ used in our simulation means that this bound is not too far below zero.

\subsection*{Connectivity learned by SAILnet allows for further experimental tests of the model}

Several previous studies of sparse coding models~\cite{olshausen96,rehn_sommer,foldiak90,falconbridge,perrinet,warland} have focused on the receptive fields learned by adaptation to naturalistic inputs, but we are aware of only one published study~\cite{garrigues} that investigated the connectivity in sparse coding models, albeit with a model that lacked biological realism. One previous study~\cite{koulakov} investigated synaptic mechanisms that could give rise to the measured distribution of connection strengths, but this work was not performed in the context of a sensory coding model. No prior work has studied the connectivity learned in a biophysically well-motivated sensory coding network, which would provide additional testable predictions for physiology experiments. 

Fig.~6 shows the distribution of non-zero connection strengths (non-zero elements of the matrix $W_{im}$) learned by a 1536-unit SAILnet with $p=0.05$ trained on $16 \times 16$ pixel patches drawn from whitened natural images (the same network whose receptive fields are shown in Fig.~3). When trained on natural images, SAILnet learns an approximately lognormal distribution of inhibitory connection strengths; a Gaussian best fit to the histogram of the logarithms of the connection strengths accounts for $98 \%$ of the variance in the data. 

Despite this close agreement, SAILnet shows some systematic deviations from the lognormal fit, especially on the low-connection-strength tail of the distribution. Interestingly, the experimental data~\cite{song} show an approximately lognormal distribution of excitatory connection strengths, with similar systematic deviations (Fig. 5b of Song \emph{et al.}~\cite{song}). By contrast, prior theoretical work~\cite{koulakov,song} has employed learning rules tailored to create exactly lognormal connection strength distributions, and thus show no such deviations. 
Note also that neither of these previous studies addressed the issue of how neurons might represent sensory inputs, nor how they might learn those representations.

Whereas the experimental data of Song \emph{et al.}~\cite{song} show a roughly lognormal distribution in the strengths of excitatory connections between V1 neurons, our model makes predictions about the strengths of \emph{inhibitory} connections in V1. The $1~ms$ time window for measuring post-synaptic potentials in the experiment of Song \emph{et al.}~\cite{song} ensured that they measured only direct synaptic connections. However, suppressive interactions between excitatory neurons in cortex are mediated by inhibitory interneurons. Consequently, the inhibitory interactions between pairs of excitatory neurons in V1 must involve two or more synaptic connections between the cells. Thus, our model predicts that the inhibitory functional connections between excitatory simple cells in V1, like the excitatory connections measured by Song \emph{et al.}~\cite{song}, should follow an approximately lognormal distribution (Fig.~6), but it does not specify the extent to which this is achieved through variations in strength among dendritic or axonal synaptic connections of V1 inhibitory interneurons. 
One recent theoretical study~\cite{clopath} has uncovered some interesting relationships between coding schemes and connectivity in cortex, but it did not make any statements about the anticipated distribution of inhibitory connections.

Interestingly, there is a clear correlation between the strengths of the inhibitory connection between  pairs of SAILnet neurons, and the overlap (measured by vector dot product) between their receptive fields: neurons with significantly overlapping receptive fields tend to have strong inhibitory connections between them (Fig.~6). This correlation is expected because cells with similar RF's receive much common feed-forward input. Thus, in order to keep their activities uncorrelated, significant mutual inhibition is required. This same feature was assumed by the LCA algorithm of Rozell~\cite{rozell} and colleagues, but is naturally learned by SAILnet, in response to natural stimuli.

Our connectivity predictions are amenable to direct experimental testing, although that testing may be challenging, owing to the difficulty of measuring functional connectivity mediated by two or more synaptic connections between pairs of V1 excitatory simple cells. 

\section{Discussion}

The present work represents the first demonstration that synaptically local plasticity rules can be used to learn a sparse code for natural images that accounts for the diverse shapes of V1 simple cell receptive fields.
Our model uses purely synaptically local learning rules --- connection strengths are updated based only on the number of spikes arriving at the synapse and the number of spikes generated by the post-synaptic cell. By contrast, the local competition algorithm (LCA) of Rozell and colleagues~\cite{rozell}
assumes that $W_{i m} = \sum_k Q_{i k} Q_{m k}$, so that the strength of the inhibitory connection between two neurons is equal to the overlap (\emph{i.e.}, vector dot product) between their receptive fields. This non-local rule requires that individual inhibitory synapses must somehow keep track of the changes in the receptive fields of many neurons throughout the network in order to update their strengths. Moreover, the LCA network does not contain spiking units, even though cortical neurons are known to communicate via discrete, indistinguishable, spikes of activity~\cite{dayan}. 

Similarly, the units in the networks of Falconbridge \emph{et al.}~\cite{falconbridge} and  F\"{o}ldi\'{a}k~\cite{foldiak90} 
communicate via continuous-valued functions of time. Although these two models~\cite{foldiak90,falconbridge} do 
use synaptically local plasticity rules, neither of these groups demonstrated that such local plasticity rules are 
sufficient to explain the diversity of simple cell RF shapes observed in V1. 

We note that, independent of the present work, Rozell and Shapero have recently implemented a spiking version of 
LCA~\cite{rozell} that uses leaky integrate-and-fire units (S. Shapero, D. Br\"{u}derle, P. Hasler, and C. Rozell, CoSyne 2011 abstract). However, that work does not address the issue of how to train such a network using synaptically local plasticity rules. 

Some groups have used spiking units to perform image coding~\cite{perrinet,perrinet_homeo,vanrullen,masquellier,delorme}, but those studies did not address the question of whether synaptically local plasticity rules can account for the observed diversity of V1 RF shapes. Interestingly, it has been demonstrated~\cite{delorme} that orientation selectivity can arise from spike timing dependent plasticity rules applied to natural scenes.
Previous work~\cite{perrinet_homeo} has also explored the addition of homeostatic mechanisms to sparse coding algorithms and found it to improve the rate at which learning converges and to qualitatively affect the shapes of the learned RFs; homeostasis is enforced in our model via modifiable firing thresholds.

Finally, we note that one previous group~\cite{savin} has demonstrated that independent component analysis (ICA) can be implemented with spiking neurons and local plasticity rules. That work did not, however, account for the diverse shapes of V1 receptive fields, although they did also demonstrate that homeostasis (a mean firing rate constraint) was critical to the learning process.

Our model attempts to be biophysically realistic, but it is not a perfect model of visual cortex in all of its details. In particular, like many previous models~\cite{olshausen96,rehn_sommer,perrinet,perrinet_homeo,warland}, our network alternates between brief periods of inference (the representation of the input by a specific population activity pattern in the network) and learning (the modification of synaptic strengths), which may not be realistic. Indeed, it is unclear how cortical neurons would ``know" when the inference period is over and when the learning period should begin, though it is interesting to note that these iterations could be tied to the onset of saccades, given the $5 \tau_{RC} \approx 100~ms$ inference period between ``learning" stages in our model.

As in previous models, the inputs to our network $X_j$ are continuous-valued, whereas the actual inputs from the lateral geniculate nucleus to primary visual cortex (V1) are spiking. As mentioned above, suppressive interactions between pairs of units in our model are mediated by direct, one-way, inhibitory synaptic connections between units, rather than being mediated by a distinct population of inhibitory interneurons. We do not include the effects of spike-timing dependent plasticity~\cite{stdp}, although this has been shown to have interesting theoretical implications for cortex~\cite{clopath} in general and for image coding in particular~\cite{masquellier,delorme}. We are currently developing models that incorporate spike timing dependent learning rules, applied to time-varying image stimuli such as natural movies.

Finally, the neurons in our model have no intrinsic noise in their activities, although that noise may, in practice, be small~\cite{mainen}.

Interestingly, since our model neurons require a finite amont of time to update their internal variables $u_i(t)$, there is a hysteresis effect if one presents the network with time-varying image stimuli --- the content of previous frames affects how the network processes and represents the current frame. Even if the features in a movie change slowly, the optimal representation of one frame can be very different from the optimal representation of the next frame in many coding models, so this hysteresis effect can provide stability to the image representation compared to other models such as ICA~\cite{bell_sejno, hyvarinen} or Olshausen and Field's sparsenet~\cite{olshausen96}. This effect has previously been studied by Rozell and colleagues~\cite{rozell}, encouraging our efforts to apply SAILnet to dynamic stimuli.

Though it is highly simplified, our model does captures many qualitative features of V1, such as inhibitory lateral connections~\cite{haider}, largely uncorrelated neuronal activities~\cite{ecker, renart}, sparse neuronal activity~\cite{haider, vinje,vinje2}, a greater number of cortical neurons than input neurons (over-complete representation), synaptically local learning rules, and spiking neurons. 
Importantly, this model allows us to make several falsifiable experimental predictions about interneuronal connectivity and population activity in cortex. We hope that these predictions will help uncover the coding principles at work in the visual cortex.

\section{Methods}

\subsection*{SAILnet dynamics}

Each of the neurons in our SAILnet follows leaky integrate-and-fire dynamics~\cite{dayan}. The neurons, indexed by subscript $i$, each have a time-dependent, continuous-valued internal variable $u_i(t)$, analogous to a neuronal membrane potential. We explicitly model each neuron as an RC circuit (Fig.~1), where the internal variable $u_i(t)$ corresponds to the voltage across the capacitor. Whenever this internal variable exceeds a threshold value $\theta_i$ specific to that neuron, the neuron emits a punctate spike of activity. The unit's external variable $y_i(t)$, which represents the spiking output that is communicated to other neurons throughout the network, is $1$ for a brief moment. At all other times, the unit's external variable is $0$.

Since the thresholds $\theta_i$ are adapted slowly compared to the time scale of inference, they are approximately constant during inference. The same is true for the feed-forward weights $Q_{ik}$ and the lateral connection strengths $W_{im}$, discussed below.

We model the effects of the input image $\{X_k\}$ and the activities of other neurons in the network $y_m(t)$ on the internal variable as a current, $I_{input}(t) = \sum_k Q_{i k} X_k - \sum_{m \ne i} W_{i m} y_{m}(t)$, that is impinging on the RC circuit; here the feed-forward weights $Q_{ik}$ and lateral connection strengths $W_{im}$ describe how much a given input (either an image pixel value, or a spike from another neuron in the network) should modify the neuron's internal variable. The internal variable evolves in time via the differential equation for voltage across our capacitor, in response to the input current $du_i(t)/dt + u_i(t) = I_{input} (t)$.

We simulate these dynamics in discrete time, performing numerical integration of the differential equation $du_i(t)/dt + u_i(t) = I_{input} (t)$. Whenever the internal variable $u_i(t)$ exceeds the threshold (at time $t^{\ast}$; $u_i(t^{\ast}) > \theta_i$), the output spike occurs at the next time step: $y_i(t^{\ast}+1) = 1$. In the subsequent time step, the external variable $y_i(t^{\ast} + 2)$ returns to zero, unless the internal variable $u_i(t)$ has again crossed the threshold.

After the unit spikes, the internal variable returns to its resting value of $0$, from whence the unit can again be charged up.

For simplicity, our differential equation assumes that the RC time constant of the model neuron is one ``unit" of time. Our simulated dynamics are allowed to run for five such units of time (with the time step of numerical integration being 0.1 units in duration), in response to each input image. At the start of these dynamics, the internal variables of all neurons are set to their resting values: $u_i(t=0)=0 ~\forall ~i$.

\subsection*{SAILnet learning rules can be viewed as a gradient descent approach to a constrained optimization problem}

Unlike previous work~\cite{olshausen96,rehn_sommer}, which performed unconstrained optimization on a cost function penalizing both reconstruction error and network activity, our learning rules can be viewed as a gradient descent approach to a \emph{constrained} optimization problem. 

Given the neuronal activities $n_i$ in response to an image, and their feed-forward weights $Q_{ik}$, one can form a linear generative model $\overline{X}$ of the input stimulus $\overline{X_k} = \sum_i n_i Q_{ik}$. The mean squared error between that model $\overline{X}$ and the true input $X$ is $E = \sum_k \left(  X_k - \sum_i n_i Q_{ik} \right) ^2$, and the creation of a high fidelity representation suggests that this error function $E$, or one like it, be minimized by the learning process.

Let us suppose that the neuronal network is not free to choose any solution to this problem; instead it must satisfy constraints that require the neurons to have a fixed average firing rate of $p$ and minimal correlation between neurons.  Indeed, neurons tend to have low mean firing rates when averaged across many different images,  and those firing rates span a finite range of values~\cite{hromadka,baddeley,abeles}, motivating our first constraint. The second constraint is justified by observations that neural systems tend to exhibit little or no correlation between pairs of units~\cite{ecker, renart}, and that the correlation between the activity of V1 neurons decreases significantly as one increases the fraction of the visual field that is stimulated~\cite{vinje2}.

We use the method of Lagrange multipliers to solve this problem, allowing our learning rules to adapt the network so as to minimize reconstruction error while approximately satisfying these constraints. To do this, we perform gradient descent on a Lagrange function $\mathcal{L}$ that contains both the error function and the constraints:

\begin{eqnarray}
\mathcal{L} &= & \sum_k \left(  X_k - \sum_i n_i Q_{ik} \right) ^2  \\
 &+& \sum_i \lambda_i (n_i-p) + \sum_{i \ne k} \tau_{im}  (n_i n_m - p^2), \nonumber
\label{eq:lagrange}
\end{eqnarray}

where the sets of values $\{\lambda_i \}$ and $\{\tau_{im} \}$ are our (unknown) Lagrange multipliers. To perform constrained optimization, gradient descent is performed with respect to all of the free parameters in $\mathcal{L}$: namely, the set of feed-forward weights $\{Q_{ik}\}$, and the Lagrange multipliers $\{\lambda_i\}$ and $\{\tau_{im}\}$:

\begin{eqnarray}
\frac{\partial \mathcal{L}}{\partial \lambda_i} & =& n_i - p  \\
\frac{\partial \mathcal{L}}{\partial \tau_{im}} &  =&  n_i n_m - p^2 \nonumber \\
\frac{\partial \mathcal{L}}{\partial Q_{ik}}  &=& -2 n_i(X_k - \sum_r n_r Q_{rk}). \nonumber 
\label{eq:descent}
\end{eqnarray}

The first two equations lead to our learning rules for inhibitory connections and firing thresholds, once we identify $\lambda_i \propto -\theta_i$ and $\tau_{im} \propto -W_{im}$;  these network parameters correspond to the Lagrange multipliers of the constrained optimization problem. This reflects the fact that the role of the variable thresholds and inhibitory connections is to enforce the sparseness and non-correlation constraints in the network, which is the same as the role of the Lagrange multipliers in the Lagrange function.

We emphasize that the terms of our objective function that effectively enforce these constraints are critical for our algorithm's success. By contrast, consider the situation in which the model units had no other possibility but to maintain their fixed firing rate and lack of correlation, due to some clever parameterization of the model's state space. In that case, one could simply minimize the reconstruction error, via gradient descent, and the existence of these extra terms, or even of the analogous  Lagrange multipliers, would be redundant. However, in our model, each change of the feed-forward weights ($Q_{ik}$) could change the neuron's firing rate, and the correlation between its activity and those of other neurons, unless something forces the network back towards the constraint surface. The variable firing thresholds and inhibitory inter-neuronal connection strengths in our model perform this function.

The last equation from our gradient descent calculation gives the update rule for the feed-forward weights $\Delta Q_{ik} \propto n_i(X_k - \sum_r n_r Q_{rk})$. This rule, as written, is unacceptable for our SAILnet because we wish to interpret the strengths of connections in that network as the strengths of synaptic connections in cortex. In that case, learning at any given synapse should be accomplished using only information available locally, at that synapse. For updating connection strength $Q_{ik}$, this could include the pre-synaptic activity $X_k$, the post-synaptic activity $n_i$, and the current value of the connection strength $Q_{ik}$, but should not require information about the receptive fields of other neurons in the network, nor their activities, because it is not clear that that information is available at each synapse. Hence, the $\sum_r n_r Q_{rk}$ term that arises from gradient descent on our objective function is a problem for the biological interpretation of these learning rules. We will now show that, in the limit that the neuronal activity is sparse and uncorrelated, when averaged over several input images, the non-local gradient descent rule $\Delta Q_{ik} \propto n_i(X_k - \sum_r n_r Q_{rk})$ is approximately equivalent to a simpler rule, originally due to Oja~\cite{Oja}, that is synaptically local.

Consider the non-local update rule $\Delta Q_{ik} \propto n_i(X_k - \sum_r n_r Q_{rk})$. Expanding the polynomial, and averaging over image presentations, we find

\begin{eqnarray}
\left< \Delta Q_{ik} \right> &\propto& \left< n_i X_k \right> - \left< n^2_i Q_{ik}  \right>- \sum_{r \ne i} \left< n_i n_r Q_{rk} \right>. 
\end{eqnarray}

If the learning rate $\beta$ is small, such that the feed-forward weights change only slowly over time, then we can approximate that they are constant over some (small) number of image presentations, and take them outside of the averaging brackets;

\begin{eqnarray}
\left< \Delta Q_{ik} \right> &\sim& \left< n_i X_k \right> - \left< n^2_i  \right> Q_{ik} - \sum_{r \ne i} \left< n_i n_r \right> Q_{rk}. 
\end{eqnarray}

Now, so long as the neuronal activities are uncorrelated, and all units have the same average firing rate (recall these constraints are enforced by our Lagrange multipliers), $\left< n_i n_r \right> = \left< n_i \right> \left< n_r \right> = p^2~\forall i,r$, and thus the learning rule is

\begin{eqnarray}
\left< \Delta Q_{ik} \right> &\sim& \left< n_i X_k \right> - \left< n^2_i  \right> Q_{ik} - p^2 \sum_{r \ne i}  Q_{rk}. 
\end{eqnarray}

This last term is small compared to the first two for a few reasons. First, the neurons in the network have \emph{sparse} activity, meaning they are selective to particular image features, and thus $\left< n_i^2 \right> \gg \left< n_i \right>^2 = p^2$. This can be easily seen by that fact that we use small values for $p$, meaning that the neurons fire, on average, much less than one spike per image. The spikes, however, can only be emitted in integer numbers, so the neurons are silent in response to most image presentations, and are thus highly selective.

Furthermore, the last term, $ p^2 \sum_{r \ne i}  Q_{rk}$, involves a sum over the receptive fields of many neurons in the network. Some of the RFs will be positive for a given pixel, whereas others will be negative. These random signs mean that the sum $\sum_{r \ne i}  Q_{rk}$ tends towards zero.

Thus,  in the limit of sparse and uncorrelated neuronal activity (the limit in which our network operates), gradient descent on the error function $E$ yields approximately

\begin{eqnarray}
\left< \Delta Q_{ik} \right> &\sim& \left< n_i X_k \right> - \left< n^2_i \right> Q_{ik}, 
\end{eqnarray}
which is equivalent to the average update from Oja's implementation of Hebbian learning~\cite{Oja}, which we use for learning in SAILnet. Thus, SAILnet learns to approximately solve the same error minimization problem as did previous, non-local sparse coding algorithms~\cite{rehn_sommer,olshausen96}. 

Interestingly, our result suggests that, despite the highly non-linear way in which our model neurons' outputs (spikes $n_i$) are generated from the input, a linear decoding of the network activity should provide a good match to the input: $X_k \approx \sum_i n_i Q_{ik}$ . This linear decodability has previously been observed in physiology experiments~\cite{bialek}, as well as models designed to maximize the information rate about input stimulus conveyed by individual spiking neurons~\cite{bialekphysicaA93,deweesenetwork96}, and it is indeed a property of SAILnet.

We summarize the learning rules for SAILnet here.

\begin{eqnarray}
\Delta \theta_i &\propto& n_i - p  \\
\Delta W_{im} & \propto&  n_i n_m - p^2 \nonumber \\
\Delta Q_{ik}  &\propto&  n_i(X_k - n_i Q_{ik}) \nonumber 
\label{eq:descent}
\end{eqnarray}

The first two rules enforce the sparseness and correlation constraints, and arise from the Lagrange multipliers in our Lagrange function. The final rule drives the SAILnet representation to form a better match to the input stimulus, as it adapts to the ensemble of training images.

\subsection*{Receptive fields measured by spike-triggered average are proportional to the feed-forward weights of the neurons when the probe stimulus statistics match those of the training stimuli.}


Consider the Oja-Hebb~\cite{Oja} learning rule for the feed-forward weights in our model, 

\begin{eqnarray}
\Delta Q_{ik} \propto n_i\left( X_k - n_i Q_{ik} \right).
\end{eqnarray}

Once the learning has converged over some set of training stimuli, the feed-forward weights are, on average, no longer changing in response to repeated presentations of examples from the training set.
Thus,

\begin{eqnarray}
\left< \Delta Q_{ik} \right> \propto \left< n_i\left( X_k - n_i Q_{ik} \right) \right> = 0.
\end{eqnarray}

Expanding the middle term in this expression,
we find that

\begin{eqnarray}
 \left< n_i  X_k \right> = \left< n_i^2 Q_{ik} \right> = \left< n_i^2 \right>Q_{ik}, 
\end{eqnarray}
where the second equality occurs because the learning has converged, and thus the feed-forward weights are constant over repeated image presentations. Thus, we find that $\left< n_i  X_k \right>/\left< n_i^2 \right> = Q_{ik}$; the spike-triggered average (STA) stimulus is equivalent to the set of feed-forward weights, up to a multiplicative scaling factor that can be calculated from the spike train.

\subsection*{Training SAILnet}
We start out each simulation with all inhibitory connection strengths $W_{im}$ set to zero, all firing thresholds $\theta_i$ set to $5$, and the feed-forward weights $Q_{ik}$ initialized with Gaussian white noise. To train the network, batches~\cite{falconbridge} of 100 images with zero mean, and unit standard deviation pixel values, are presented, and the number of spikes from each neuron are counted separately for each image. After each batch, the average update for the network properties is computed (following our learning rules) over the 100-image batch. This batch-wise training lets us use matrix operations for computing the updates, which dramatically speeds up the training process. After each update, all negative values for inhibitory connections $W_{im}$ (which would correspond to excitatory connections) are set to zero, as in the previous work by F\"{o}ldi\'{a}k~\cite{foldiak90}. Relaxing this constraint, and allowing the recurrent weights to change sign does not affect our qualitative conclusions. In that case, some of the recurrent connections become excitatory, while the majority remain inhibitory, the RF's are qualitatively the same as those shown in Fig.~3, and the distributions of inhibitory and excitatory connection strengths are both approximately lognormal (data not shown).

The relative values of $\alpha,~\beta$ and $\gamma$ were chosen based on F\"{o}ldi\'{a}k's~\cite{foldiak90} observation that $\beta$ must be much less than $\alpha$ or $\gamma$ so that the neurons' activities remain sparse and uncorrelated, even in the face of changing feed-forward weights. 

We study the network after the properties stop changing macroscopically over time. However, as noted in the firing rates section of this paper, the network parameters continue to bounce around the final ``target" state, with the size of the bounces determined by the learning rates in the network. Empirically, we find that it takes on the order of $10^7$ image presentations ($10^5$ steps of 100 image presentations per step) for this dynamic equilibrium to be established. For the results presented in this paper, we let the network train for roughly $2\times 10^8$ image presentations.

To speed up the simulation, we start the training with large values for the learning rates, and these are eventually reduced. For the last $10^4$ batches of training ($10^6$ image presentations), the learning rates were $(\alpha, \beta, \gamma) = (0.1,0.001,0.01)$.

All of the computer codes used to generate the results presented in this paper are available upon request.

\subsection*{Fitting SAILnet RFs to Gabor functions}
A Gabor function ($G(x,y)$) is a common model for visual cortical receptive fields~\cite{rehn_sommer,Ringach}, which consists of a two-dimensional Gaussian multiplied by a sinusoid:

\begin{eqnarray}
G &=& A \cos(2 \pi f x_p + \psi) \exp \left[ -\left(\frac{x_p}{\sqrt{2}\sigma_x}\right)^2 - \left(\frac{y_p}{\sqrt{2}\sigma_y}\right)^2   \right],  \\
x_p &=& (x-x_0) \cos(\theta) + (y-y_0) \sin(\theta), \nonumber \\
y_p &=& -(x-x_0) \sin(\theta) + (y-y_0) \cos(\theta). \nonumber
\end{eqnarray}

The center of the shape is defined by the coordinates $x_0$ and $y_0$, while the amplitude and orientation of the pattern are defined by the parameters $A$ and $\theta$, respectively. $\psi$ defines the phase of the sinusoid, relative to the center of the Gaussian envelope, which has spatial extent $\sigma_x$ and $\sigma_y$ in the direction along, and perpendicular to, the direction in which the sinusoid oscillates (with frequency $f$), respectively. 

Given a neuronal RF, our code performs unconstrained optimization to choose the Gabor parameters such that the mean squared error $||G - RF||^2$ is minimized. We then perform several quality control measures to ensure that our analysis only contains sensible Gabor parameters that accurately describe our RFs.

The first such measure is to exclude any RF for which the deviation between the RF and the Gabor fit is large; cells with $||G - RF||^2/||RF||^2 >0.5$ were excluded. This is equivalent to placing a (fairly mild) restriction on the minimum allowable signal-to-noise ratio.

The second quality control measure is to exclude those RFs for which the center of the pattern $(x_0,y_0)$ lies either outside the $16 \times 16$ pixel patch, or within one standard deviation (of the Gaussian envelope) of the patch edge. As described by other workers~\cite{rehn_sommer}, when the center of the pattern is outside of the visible $16\times16$ pixel patch, it is not clear that the shape of the RF itself is well-described by the Gabor parameters, or even well-constrained, for that matter. Our (more stringent) restriction also avoids the problem of biased shape estimates, when fitting Gabors to RFs that are truncated by the edge of the patch; the model RFs essentially tile the available space, so some of them will, by necessity, have centers that lie right along or outside of the edges of the patch. Indeed, in a $16 \times 16$ pixel space, many pixels are near the edge, thus this cut excludes many RFs.

After making all of these cuts, we were left with 299 RFs, on which to perform subsequent shape analysis. 

We performed the same fitting and quality-control analysis on both the SAILnet and the macaque physiology RFs, although we used a gentler goodness-of-fit restriction on the macaque data, since the macaque RFs, as measured, have fairly large regions of zero support, in which any measurement noise reduces the apparent goodness-of-fit. For the macaque data, we excluded those RFs with  $||G - RF||^2/||RF||^2 >0.8$, leaving 116 of the 250 macaque RFs for subsequent analysis.

One reason for the relatively low yield of well-fit RFs is that not all RFs are actually well-described by Gabor functions. For example, there is no choice of Gabor parameters that will accurately describe a center-surround receptive field; that RF is much better described by a difference of Gaussians function, for example. We leave for future work the issue of determining the best family of functions with which to describe visual cortical receptive fields.



%
%

%



\section*{Acknowledgments}
The authors are grateful to Bruno Olshausen for his role in inspiring this work and to Bruno Olshausen, Jascha Sohl-Dickstein, Fritz Sommer, and the other members of the Redwood Center for many helpful discussions. We are also very grateful to Dario Ringach for providing the macaque physiology data, and to Chris Rozell and Jascha Sohl-Dickstein for constructive comments on the manuscript. The work of JZ is supported by funds from the University of California and a Fulbright international science and technology PhD fellowship. MRD gratefully acknowledges support from the National Science Foundation, the McKnight Foundation, the McDonnell Foundation, and the Hellman Family Faculty Fund.

\section*{Author Contributions}
JZ initiated the project, designed and implemented the simulations, and performed the analytic calculations. JTM implemented the Gabor fitting routine. JZ performed the remainder of the data analysis. JZ and MRD wrote the manuscript. 

\section*{Additional Information}
\subsubsection*{Competing financial interests} The authors declare no competing financial interests.

 \begin{figure}[h!]
 \includegraphics[width=3.0in]{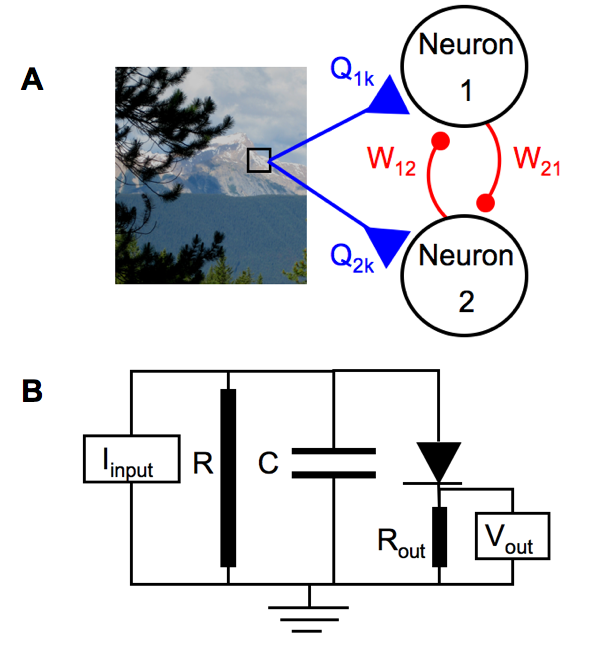}
 \caption{{\bf SAILnet network architecture and neuron model} (A) Our network architecture is based on those of Rozell \emph{et al.}~\cite{rozell} and F\"{o}ldi\'{a}k~\cite{foldiak90,falconbridge}, and inspired by recent physiology experiments~\cite{haider,vinje,ecker}. Inputs $X_k$ to the network (from image pixels) contact the neuron at connections (synapses) with strengths $Q_{ik}$, whereas inhibitory recurrent connections between neurons~\cite{haider} in the network have strengths $W_{im}$. The outputs of the neurons are given by $y_i(t)$; these spiking outputs are communicated through the recurrent connections, and also on to subsequent stages of sensory processing, such as cortical area V2, which we do not include in our model. (B) Circuit diagram of our simplified leaky integrate-and-fire~\cite{dayan} neuron model. The inputs from the stimulus with pixel values $X_k$, and the other neurons in the network, combine to form the input current $I_{input}(t) = \sum_k Q_{i k} X_k - \sum_{m \ne i} W_{i m} y_{m}(t)$ to the cell. This current charges up the capacitor, while some current can leak to ground through a resistor in parallel with the capacitor. The resistors are shown as cylinders to highlight the fact that they model the collective action of ion channels in the cell membrane. The internal variable evolves in time via the differential equation for voltage across our capacitor, in response to input current $I_{input}$: $du_i(t)/dt + u_i(t) = I_{input} (t)$, which we simulate in discrete time. Once that voltage exceeds threshold $\theta_i$, the diode, which models neuronal voltage-gated ion channels, opens, causing the cell to fire a punctate action potential, or spike, of activity. For sake of a complete circuit diagram, the output is denoted as the voltage, $V_{out}$, across some (small: $R_{out} \ll R$) resistance. After spiking, the unit's internal variable returns to the resting value of $0$, from whence it can again be charged up.}
 \end{figure}

 \begin{figure}[h!]
 \includegraphics[width=6.0in]{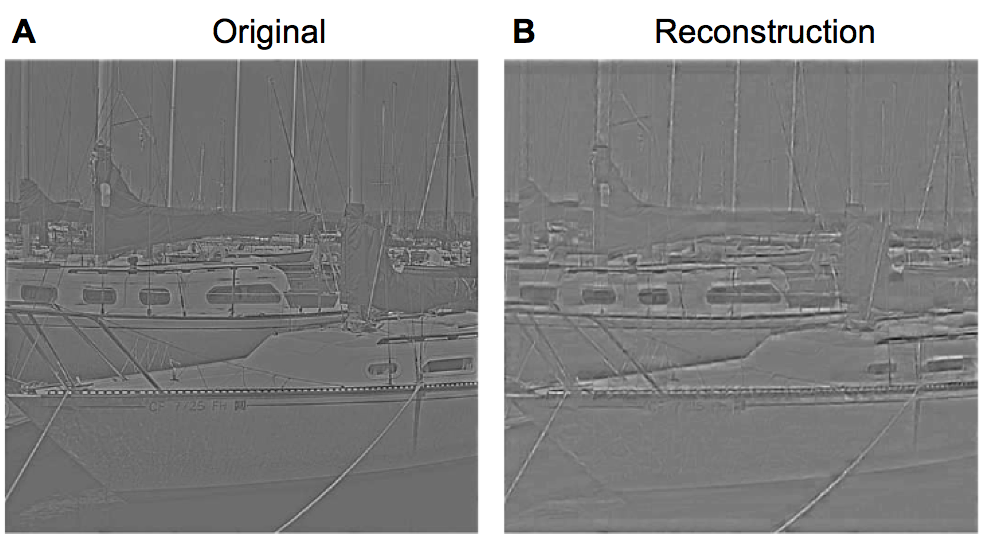}
 \caption{{\bf SAILnet activity can be linearly decoded to approximately recover the input stimulus}
(A) An example of an image that was whitened using the filter of Olshausen and Field~\cite{olshausen96}, which is the same filter used to process the images in the training set. The image in panel (A) was not included in the training set. (B) A reconstruction of the whitened image in (A), by linear decoding of the firing rates of SAILnet neurons, which were trained on a different set of natural images. The input image was divided into non-overlapping $16 \times 16$ pixel patches, each of which was preprocessed so as to have zero-mean and unit variance of the pixel values (like the training set). Each patch was presented to SAILnet, and the number of spikes were recorded from each unit in response to each patch. A linear decoding of SAILnet activity for each patch $\overline{X_k} = \sum_i n_i Q_{ik}$ was formed by multiplying each unit's activity by that unit's RF and summing over all neurons. The preprocessing was then inverted, and the patches were tiled together to form the image in panel (B). The decoded image resembles the original, but is not identical, owing to the severe compression ratio; on average, each $16 \times 16$ input patch, which is defined by 256 continuous-valued parameters, is represented by only 75 binary spikes of activity, emitted by a small subset of the neural population. Linear decodability is a product of our learning rules, and it is an observed feature of multiple sensory systems~\cite{bialek} and spiking neuron models optimized to maximize information transmission~\cite{bialekphysicaA93,deweesenetwork96}.}
 \end{figure}

 \begin{figure}[h!]
 \includegraphics[width=5.5in]{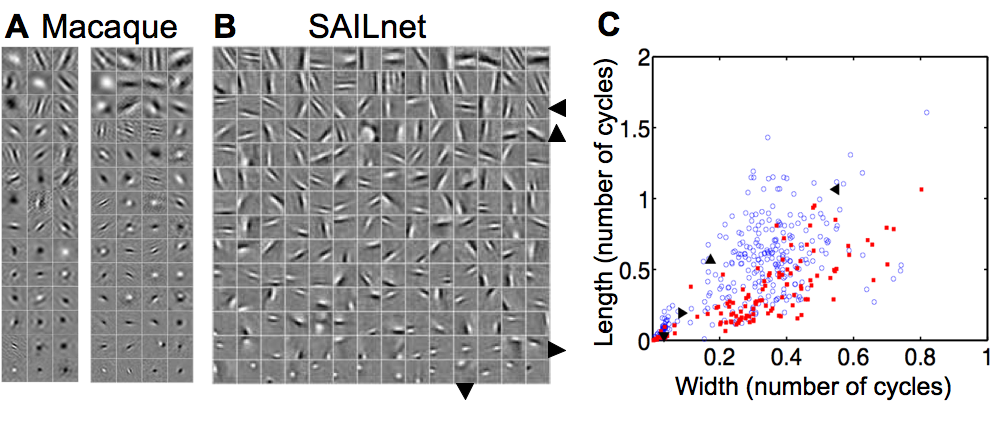}
 \caption{{\bf SAILnet learns receptive fields (RFs) with the same diversity of shapes as those of simple cells in macaque primary visual cortex (V1)} (A) 98 randomly selected receptive fields recorded from simple cells in macaque monkey V1 (courtesy of D. Ringach). Each square in the grid represents one neuronal RF. The sizes of these RFs, and their positions within the windows, have no meaning in comparison to the SAILnet data. The data to the right of the break line have an angular scale (degrees of visual angle spanned horizontally by the displayed RF window) 
 of $0.94^{o}$, whereas those to the left of it span $1.88^{o}$. (B) RFs of 196 randomly selected model neurons from a 1536-unit SAILnet trained on patches drawn from whitened natural images. The gray value in all squares represents zero, whereas the lighter pixels correspond to positive values, and the darker pixels correspond to negative values. All RFs are sorted by a size parameter, determined by a Gabor function best fit to the RF. The SAILnet model RFs show the same diversity of shapes as do the RFs of simple cells in macaque monkey V1 (A); both the model units and the population of recorded V1 neurons consist of small unoriented features, oriented Gabor-like wavelets containing multiple subfields, and elongated edge-detectors. (C) We fit the SAILnet and macaque RFs to Gabor functions (see Methods section), in order to quantify their shapes. Shown are the dimensionless \emph{width} and \emph{length}  parameters ($\sigma_x \times f$ and $\sigma_y \times f$, respectively) of the 299 SAILnet RFs and 116 (out of 250 RFs in the dataset) macaque RFs for which the Gabor fitting routine converged. These parameters represent the size of the Gaussian envelope in either direction, in terms of the number of cycles of the sinusoid. The SAILnet data (open blue circles) span the space of the macaque data (solid red squares) from our Gabor fitting analysis; SAILnet is accounting for all of the observed RF shapes. We highlight four SAILnet RFs with distinct shapes, which are identified by the large triangular symbols that are also displayed next to the corresponding RFs in panel (B). }
 \end{figure}

\begin{figure}[ht!]
\includegraphics[width=5.5in]{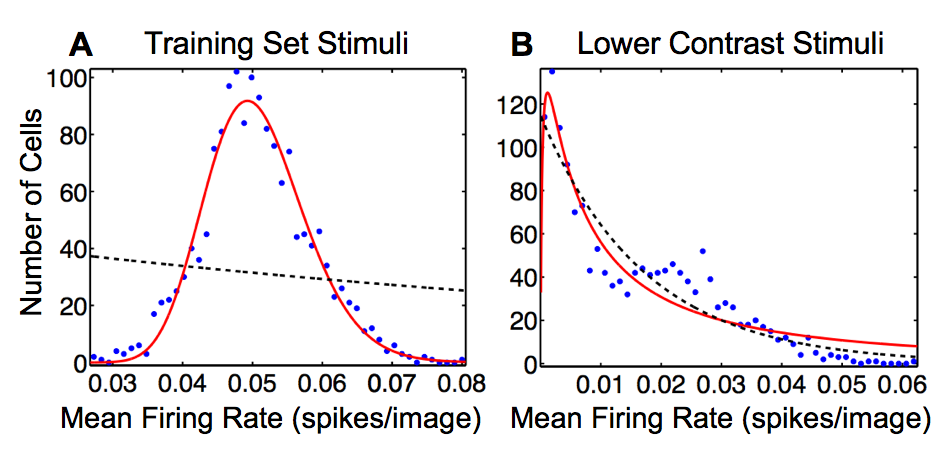}
\caption{{\bf Units in SAILnet exhibit a broad range of mean firing rates, which can be lognormally or exponentially distributed depending on the choice of probe stimuli.} (A) Frequency histogram of firing rates averaged over $50,000$ image patches drawn from the training ensemble for each of the 1536 units of a SAILnet trained on whitened natural images. All learning rates were set to zero during probe stimulus presentation. A wide range of mean rates was observed, but as expected, the distribution is peaked near $p = 0.05$ spikes per image, the target mean firing rate of the neurons. The paucity of units with near-zero firing rates suggests that this distribution is closer to lognormal than exponential. Accordingly, the lognormal least-squares (solid red curve) fit accounts for $R^2 = 96~\%$ of the variance in the data, whereas the exponential fit (black dashed curve) accounts for only $2~\%$. (B) In response to low contrast stimuli, the firing rate distribution across the units (every unit fired at least once) in the same network as in panel (A) was similarly well fit by either an exponential (dashed black curve; accounting for $R^2 = 88~\%$ of the variance in the data) or a lognormal function (solid red curve; accounting for $90~\%$ of the variance).
The low-contrast stimulus ensemble used to probe the network consisted of images drawn from the training set, with all pixel values reduced by a factor of three.}
\end{figure}

\begin{figure}[ht]
\includegraphics[width=2.8in]{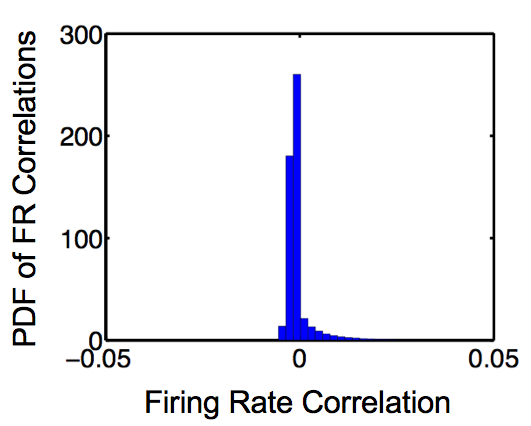}
\caption{{\bf Pairs of SAILnet units have small firing rate correlations. }
The probability distribution function (PDF) of the Pearson's linear correlation coefficients between the spike-counts of pairs of SAILnet neurons responding to an ensemble of 30,000 natural images is sharply peaked near zero.}
\end{figure}

\begin{figure}[ht]
\includegraphics[width=5.5in]{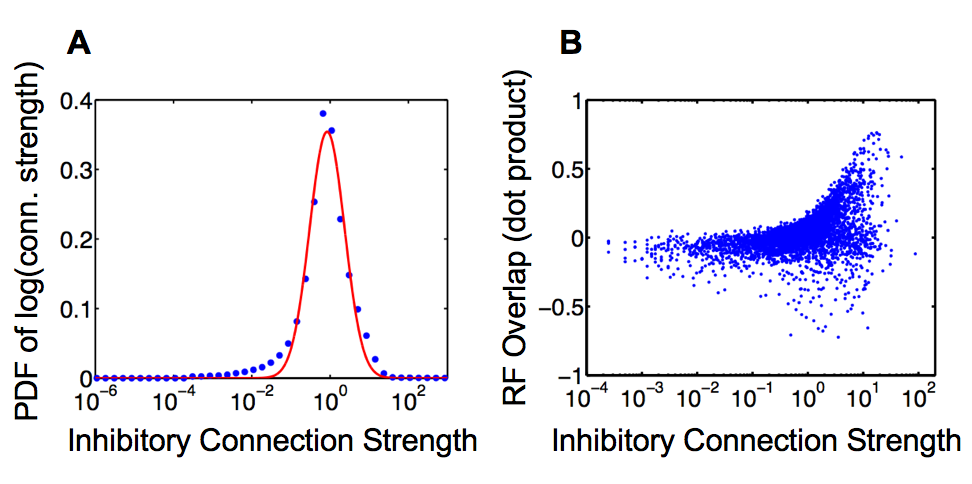}
\caption{{\bf Connectivity learned by SAILnet allows for further experimental tests of the model.}
(A) Probability Density Function (PDF) of the logarithms of the inhibitory connection strengths (non-zero elements of the matrix $W_{im}$) learned by a 1536 unit SAILnet trained on $16\times16$ pixel patches drawn from whitened natural images. The measured values (blue points) are well-described by a Gaussian distribution (solid line), which accounts for $R^2 = 98~\%$ of the variance in the dataset. This indicates that the data are approximately lognormally distributed. Note that there are some systematic deviations between the Gaussian best fit and the true distribution, particularly on the low-connection strength tail, similar to what has been observed for excitatory connections within V1~\cite{song}. This plot was created using the binning procedure of Hrom\'{a}dka and colleagues~\cite{hromadka}. The histogram was normalized to have unit area under the curve. (B) The strengths of the inhibitory connections between pairs of cells are correlated with the overlap between those cells' receptive fields: cells with significantly overlapping RFs tend to have strong mutual inhibition. Data shown in panel (B) are for 5,000 randomly selected pairs of cells. Pairs of cells with significantly negatively overlapping RFs tend not to have inhibitory connections between them, hence the apparent asymmetry in the RF overlap distribution obtained by marginalizing over connection strengths in panel (B).}
\end{figure}

\end{document}